\documentclass{article}
\usepackage{spconf,amsmath,graphicx}

\title{The sound of my voice:\\
Speaker representation loss for target voice separation}

\name{Seongkyu Mun*, Soyeon Choe*\thanks{\hspace{-12pt}* These authors contributed equally to this work.}, Jaesung Huh, Joon Son Chung}
\address{Naver Corporation}
\address{Naver Corporation, South Korea}

\begin{document}
%
\maketitle
\begin{abstract}
Content and style representations have been widely studied in the field of style transfer. In this paper, we propose a new loss function using speaker content representation for audio source separation, and we call it speaker representation loss. The objective is to extract the target speaker voice from the noisy input and also remove it from the residual components. Compared to the conventional spectral reconstruction, our proposed framework maximizes the use of target speaker information by minimizing the distance between the speaker representations of reference and source separation output. We also propose triplet speaker representation loss as an additional criterion to remove the target speaker information from residual spectrogram output. VoiceFilter framework is adopted to evaluate source separation performance using the VCTK database, and we achieved improved performances compared to the baseline loss function without any additional network parameters.
\end{abstract}
\section{Introduction}
\label{sec:intro}

A number of recent studies have used neural network for modeling perception. Based on the remarkable performance improvement in visual object classification~\cite{wu2019wider}, neural network perception has been utilized not only for classification task, but also for training image generation~\cite{salimans2016improved} and style transfer~\cite{gatys2016image, jamaludin2019you, chen2017photographic}. These methods use the high-level image representations of neural networks instead of measuring the distance or distribution between the raw image pixels. For example, images of black and white car would have a huge difference at the pixel level, whereas they would show a high similarity in terms of a “content representation” of object recognition network~\cite{gatys2016image, chen2017photographic}. Having this concept in mind, this approach would certainly be applicable to audio, also.

An interesting paper~\cite{grinstein2018audio} has been released lately, showing sound generation using image style transfer techniques. This network used statistical approach to extract the stationary sound texture, and the style of sound was transferred to human-like voice or other musical instruments. Seeing the possibility of adapting image style transfer to audio, there has not yet been an approach using “content representation” for loss function in audio pre-processing field.

Particularly, audio source separation is a suitable topic to adopt content representation in the training of pre-processing. Recent neural network-based approaches, \textit{e.g.} deep clustering~\cite{hershey2016icassp}, deep attractor network (DANet)~\cite{chen2017icassp} and permutation invariant training (PIT)~\cite{yu2017icassp, kolbaek2017taslp}, already use neural network embedding to output separated voices.

A related research named \textit{VoiceFilter}~\cite{wang2019interspeech} (VF) was published recently, for targeting speaker-dependent speech separation \cite{delcroix2018single,vzmolikova2017learning,wang2018deep}. To separate the target voice among multiple speakers, VF uses target speaker embedding as an enrollment feature, which is extracted from a pre-trained and fixed speaker recognition network. By having this enrollment stage, VF overcomes the ambiguity in previous speech separation methods~\cite{hershey2016icassp, chen2017icassp, kolbaek2017taslp} of selecting the target speaker among separated outputs. Adopting a widely used model in speaker recognition field~\cite{chung2018interspeech, nagrani2017interspeech}, we propose to use the speaker embedding as content representation. 

\begin{figure}[!t]
     \centering
     \includegraphics[width=\linewidth]{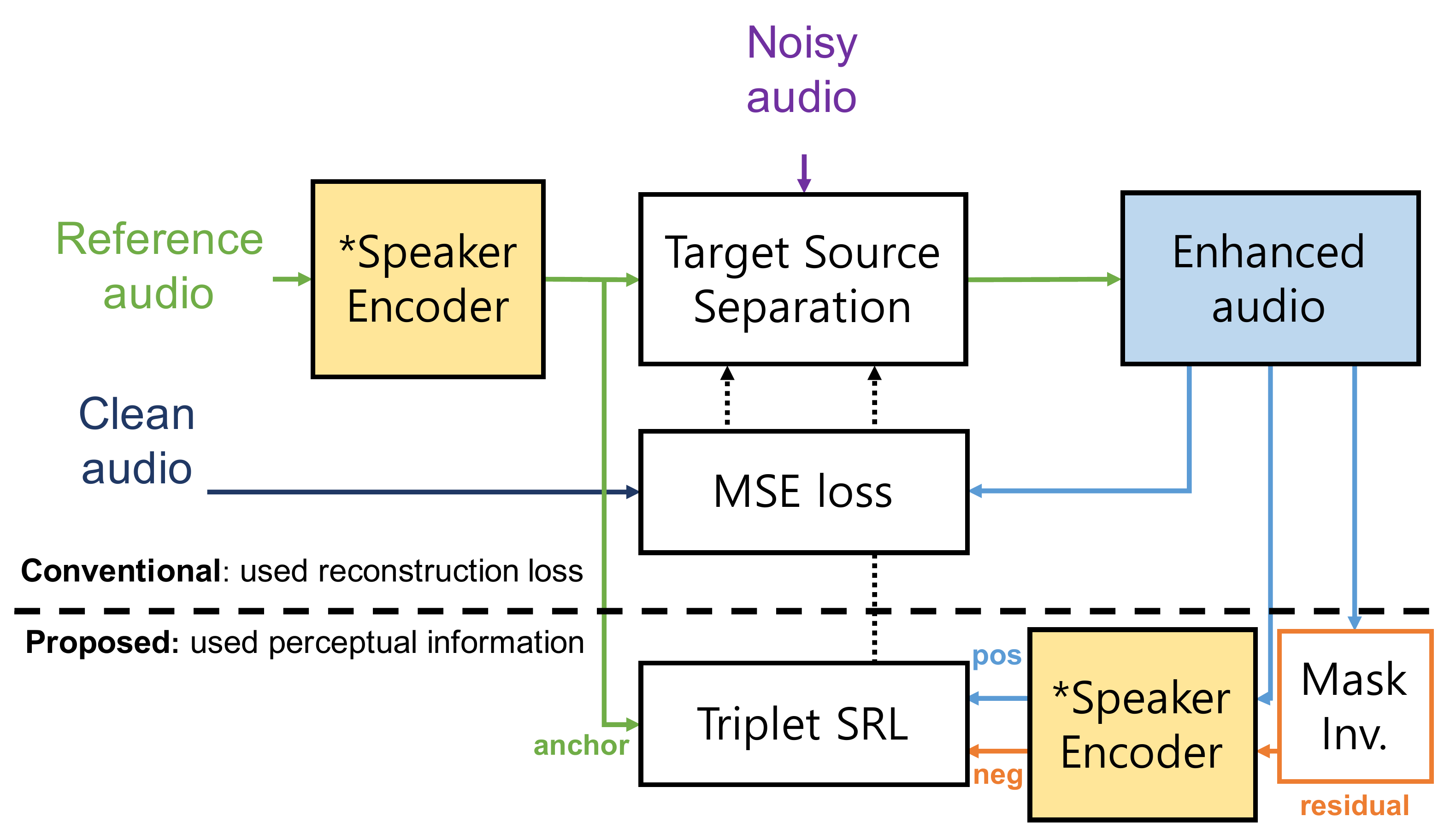}
     \caption{The concept of proposed triplet SRL. All speaker encoders indicate the same pre-trained model.}
    \label{fig:1}
\end{figure}

We extend the conventional loss of VF in two ways. First approach is minimizing the distance between the speaker representations of the estimated output and the target speaker audio. Second approach is similar to the first one, except computing the residual output to maximize the distance between itself and the clean/reference spectrogram as shown in Fig.\ref{fig:1}. 
These proposed approaches are not only limited to VoiceFilter framework, but they can be easily applied to other neural network-based source separation systems with the presence of a pre-trained speaker network. The reason this paper uses VF as the baseline is that VF framework already utilizes speaker content representation for the input. Therefore, the proposed system is implemented in a framework similar to VF without any additional settings in the speaker network.

Rest of the paper consists of five sections as follows. Section 2 describes the details of the VF framework followed by the proposed algorithm in Section 3. Furthermore, the experimental results and conclusion are indicated in Section 4 and Section 5, respectively.

\begin{figure*}[!htb]
\centering 
\includegraphics[width=0.69\linewidth]{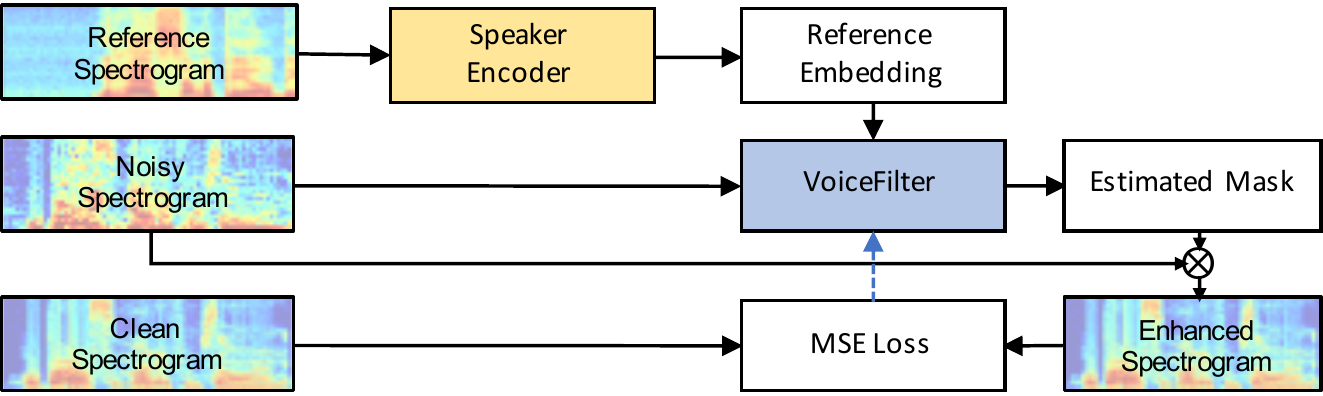}
\caption{VoiceFilter framework. Conventional VoiceFilter system architecture that outputs estimated soft mask based on reconstruction error.}
\label{fig:2} 
\end{figure*}

\begin{figure*}[!htb]
\centering 
\includegraphics[width=0.63\textheight]{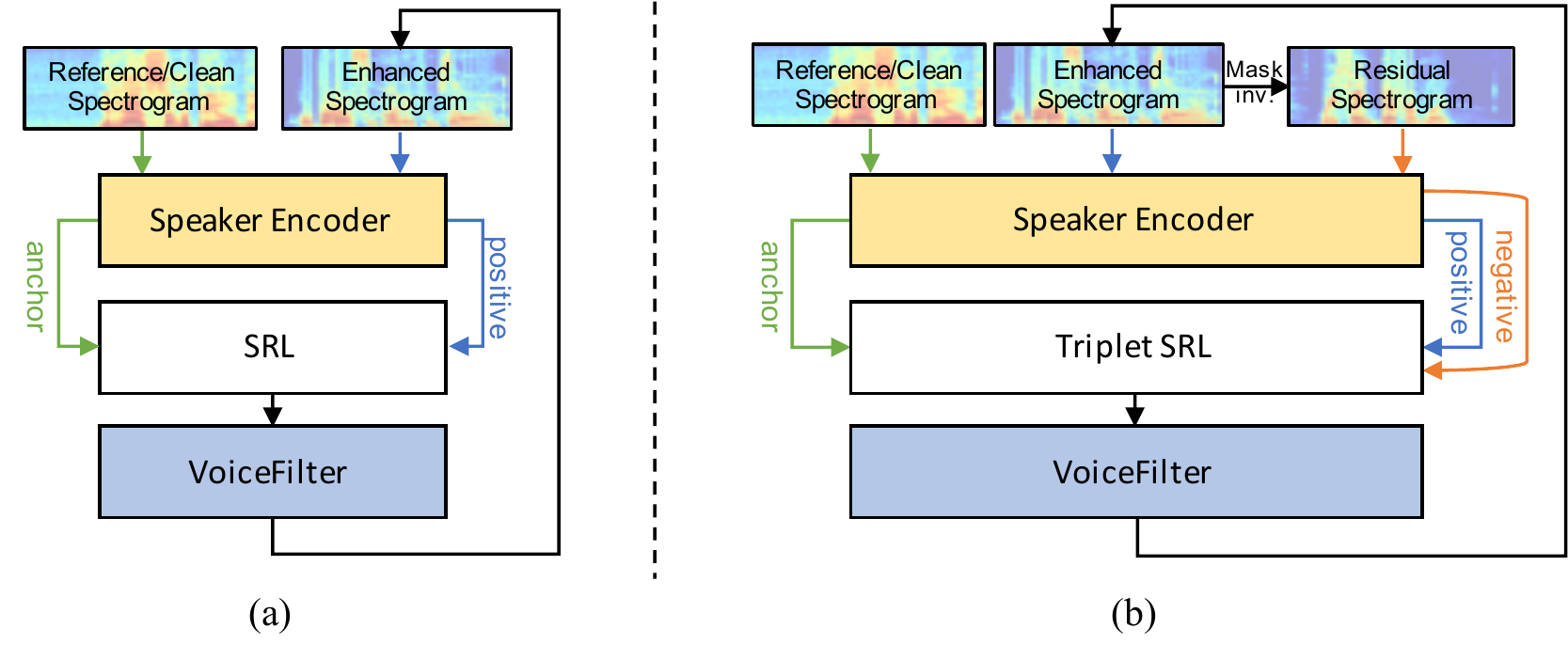}
\caption{Proposed loss function flow. (a) and (b) indicate our proposed \textit{speaker representation loss (SRL)} and \textit{triplet SRL}, which use speaker perceptual information. These are applied as \textbf{additional criteria} in VoiceFilter along with the existing reconstruction loss. }
\label{fig:3} 
\end{figure*}

\section{Related Works}
\label{sec:related_works}
\subsection{Speaker Encoder}
In order to separate the target speaker, speaker embedding, \textit{i.e.} d-vector, needs to be formed using speaker encoder. We obtained d-vector with the dimension of 512, using five convolutional layers and three fully-connected layers based on~\cite{nagrani2017interspeech}, which is different from the three-layer LSTM network indicated in VF framework. Moreover, second to last fully-connected layer was used to output d-vector without the activation function. We trained the model using VoxCeleb2~\cite{chung2018interspeech} training set and tested with VoxCeleb1~\cite{nagrani2017interspeech} test set, observing 6.0\% equal error rate of speaker verification performance. This network modified efficient image-based architecture to output valid results with spectrogram inputs. Spectrogram is generated using sliding hamming window of 25ms with 10ms step size for maximum 300 frames. Note that the speaker encoder weights are not updated in VF training phase.
\subsection{VoiceFilter}
VF system aims to enhance a noisy magnitude spectrogram by predicting a soft mask that resembles only the target speaker voice. Based on the work of Wang \textit{et al.}~\cite{wang2019interspeech}, soft mask is computed using a network of eight convolutional layers, one LSTM layer, and two fully-connected layers. Having both target speaker embedding, \textit{i.e.} d-vector, and noisy magnitude spectrogram as inputs, VF trains the network to minimize the loss between the enhanced magnitude spectrogram and clean magnitude spectrogram of the target speaker. For the guidance of training the target speaker voice, the d-vector is injected to the last convolutional layer for every frame.  


\subsection{Content Loss}
With the increased usage of neural network perception in evaluating training loss, the concept of ‘content loss’ was introduced in image-based architecture~\cite{gatys2016image, chen2017photographic}. Instead of measuring the distance between raw image pixels, this loss uses high-level image representations of neural networks to obtain high similarity in terms of “content representation”. 

As shown in Fig.~\ref{fig:2}, the loss function in the VF training phase is based on the reconstruction error between the clean and estimated spectrogram, which does not make any use of content representation of the output. Therefore, adding speaker content to the training loss could lead to performance improvement by the use of speaker perceptual information. Consequently, we utilized the speaker network not only for input conditioning (guiding target speaker), but also for verification that the target speaker is the only voice present in the output signal of the VF.

\section{Proposed Approach}
\label{sec:approach}
 
We propose a \textit{speaker representation loss} (SRL), which trains to minimize the distance between the speaker representation of the target speaker audio and the estimated output. This proposed loss is defined as a pairwise-distance between embedding vectors of two spectrograms extracted using the speaker encoder network. The embedding vectors are L2 normalized individually, and the Euclidean distance is selected for distance measure through iterative experiments.

\subsection{Anchor Embedding}
As shown in Fig.~\ref{fig:3} (a), speaker contents extracted from the masked and reference spectrograms are appended to the conventional VF. Fig.~\ref{fig:3} (a) and (b) indicate that the anchor can be selected in two ways. First approach is setting the reference embedding vector as an anchor, since VF aims to extract the output voice corresponding to the speaker of the reference audio. Second approach is setting the clean embedding vector in the training phase as an anchor, which uses the same source information with the spectrogram reconstruction. Both approaches use the same speaker's utterances as the anchor. Furthermore, since we mix the clean audio with other interfering voices, the second approach can guide the network to learn the target speaker's information more precisely.

\subsection{Negative Embedding}
Unlike image processing, masking-based source separation systems, including VF, can generate a residual spectrogram using estimated target speech and noisy input. Based on the complementary relation between the masked and residual spectrograms, we used embedding feature extracted from the residual spectrogram as a negative sample to reduce target speaker content. This concept is shown in Fig.~\ref{fig:3} (b). The network is trained to minimize the distance between speaker content of VF output and clean/reference spectrogram, along with maximizing the distance between residual output and clean/reference spectrogram. Ideally, the target speaker content should not be in residual spectrogram, but only in masked spectrogram. This process is implemented in the form of \textit{triplet loss}, which is widely used in metric learning-based approaches~\cite{schroff2015facenet,cheng2016person}. More details will be covered in the following sections.

\subsection{Speaker Representation Loss (SRL)}
With iterative experiments, following processes are used for the proposed VF. Let $f(\mathbf{x};\mathbf{w}) \in \mathbf{R}^D$ be the speaker encoder that maps the input to embedding space of $D$ dimensions.
With the sample index $i$, spectrograms of anchor (reference or clean), positive (enhanced) and negative samples (residual) are indicated as $\mathbf{x}_{i}^A$, $\mathbf{x}^P_{i}$ and $\mathbf{x}^N_{i}$. $N(\mathbf{x})$ is L2 normalization function for embedding vector, $d[\mathbf{x} , \mathbf{y}]$  is the Euclidean distance between $\mathbf{x}$ and $\mathbf{y}$, and $L_{mse}$ is mean squared error (MSE) loss function. Compared to \cite{wang2019interspeech}, we used MSE loss for spectrogram reconstruction instead of power-law compressed reconstruction error for simplicity.

We conducted experiments using SRL between anchor and positive pair. For each pair $P_{i}^+ = (\mathbf{x}_{i}^A, \mathbf{x}^P_{i})$, the overall loss $L$ combines the conventional MSE loss and the speaker representation distance ($D_{SR}$) with positive pair to be minimized per batch:

\begin{flalign}
\label{eq:PosLos}
    \nonumber &L {} =  \sum_{i} (\;\;\beta * D_{SR}(P_{i}^+;\mathbf{w}) + L_{mse}(\mathbf{x}_{i}^{clean},\mathbf{x}_{i}^P)\;\;)&&\\
    &D_{SR}(P_{i}^+;\mathbf{w})  = { { d[\; N(f(\mathbf{x}_{i}^A)) ,\; N(f(\mathbf{x}_{i}^P))\; ]}} ,&&  
\end{flalign}\\
where $\mathbf{x}_{i}^{clean}$ is a clean spectrogram for training and $\beta$ is a constant weight (\textit{e.g.} $\beta = 0.3$). If  the $\beta$ is  set  to 0,  the  loss  function  would be  the  same  as  the  original  VF.

\subsection{Triplet SRL}
Triplet SRL is a loss function that minimizes the distance between the embedding vectors from the same speaker (anchor and positive samples) and maximizes the distance between different speakers (anchor and negative samples), simultaneously. For each pair $P_{i}^+ = (\mathbf{x}_{i}^A, \mathbf{x}^P_{i})$ and $P_{i}^- = (\mathbf{x}_{i}^A, \mathbf{x}^N_{i})$, triplet SRL is minimized with the following equation per batch with conventional MSE loss:
\begin{flalign}
\label{eq:triplet}
    \nonumber&L {} =  \sum_{i} (\beta * L_{tri} + L_{mse}(\mathbf{x}_{i}^{clean},\mathbf{x}_{i}^P))&&\\
    \nonumber&L_{tri} = max[\;0, D_{SR}(P_{i}^+;\mathbf{w}) - D_{SR}(P_{i}^-;\mathbf{w}) + \alpha]&&\\
    \nonumber&D_{SR}(P_{i}^+;\mathbf{w})  = { { d[\; N(f(\mathbf{x}_{i}^A)),\; N(f(\mathbf{x}_{i}^P))\; ]}}&&\\
    &D_{SR}(P_{i}^-;\mathbf{w})  = { { d[\; N(f(\mathbf{x}_{i}^A)),\; N(f(\mathbf{x}_{i}^N))\; ]}} ,&&
\end{flalign}
where $\alpha$ and $\beta$ are constant margin and weight (\textit{e.g.} $\alpha = 1, \beta = 0.3$). 
Since the purpose of SRL is to improve spectral reconstruction, MSE loss is always included in the training criteria of VF. Through the SRL-based training, error feedback of the target speaker presence can be used in addition to the conventional spectrogram reconstruction error.

\section{Experiments}
\subsection{Data Generation and Settings}
As indicated in~\cite{wang2019interspeech}, we train and evaluate our network with VCTK dataset~\cite{veaux2016SUPERSEDEDC} to achieve a reliable comparison between the conventional VF and our proposed method. 99 and 10 speakers are randomly selected for each training and validation, followed by the data generation workflow also referred in~\cite{wang2019interspeech}. Parameter settings are mostly the same as the conventional VF, except for the d-vector dimension being set to 512. 

As mentioned in Section 3.1, we conduct the experiments separately for each of the two anchor selections (reference or clean audio). In the initial training iteration, training with only the reconstruction loss showed faster convergence. Therefore, SRL is used after 5 epoch steps for all experiments.

The network is trained using adaptive movement estimation (Adam) optimizer~\cite{kingma2014adam} with an initial learning rate of $10^{-3}$, decreasing by a factor of 0.99 for every epoch. The training is stopped after 150 epochs or whenever the validation error did not improve for 10 epochs, whichever is sooner.

\subsection{Evaluation and Results}

\begin{table}[]
\begin{tabular}{|l|c|c|}
\hline
\multicolumn{3}{|c|}{\begin{tabular}[c]{@{}c@{}}Noisy audio without VoiceFilter (Baseline) :\\ Mean SDR : 6.001 {[}dB{]}, PESQ : 2.059\end{tabular}}                                                                          \\ \hline
\multicolumn{1}{|c|}{(mean $\pm$ std.)}                                                   & \textbf{\begin{tabular}[c]{@{}c@{}}Mean SDR\\ improv.\end{tabular}} & \textbf{\begin{tabular}[c]{@{}c@{}}PESQ\\ improv.\end{tabular}}   \\ \hline
Conventional VoiceFilter {[}11{]}                                                     & \begin{tabular}[c]{@{}c@{}}4.617\\ $\pm0.047$\end{tabular}            & \begin{tabular}[c]{@{}c@{}}0.526\\ $\pm0.001$\end{tabular}          \\ \hline
\begin{tabular}[c]{@{}l@{}}+ SRL\\ (with reference audio anchor)\end{tabular}                & \begin{tabular}[c]{@{}c@{}}4.960\\ $\pm0.031$\end{tabular}            & \begin{tabular}[c]{@{}c@{}}0.587\\ $\pm0.002$\end{tabular}          \\ \hline
\begin{tabular}[c]{@{}l@{}}+ SRL \\ (with clean audio anchor)\end{tabular}                   & \begin{tabular}[c]{@{}c@{}}4.996\\ $\pm0.028$\end{tabular}            & \begin{tabular}[c]{@{}c@{}}0.593\\ $\pm0.002$\end{tabular}          \\ \hline
\begin{tabular}[c]{@{}l@{}}+ Triplet SRL\\ (with reference audio anchor)\end{tabular} & \begin{tabular}[c]{@{}c@{}}4.953\\ $\pm0.019$\end{tabular}             & \begin{tabular}[c]{@{}c@{}}0.592\\ $\pm0.001$\end{tabular}          \\ \hline
\begin{tabular}[c]{@{}l@{}}+ Triplet SRL\\ (with clean audio anchor)\end{tabular}     & \textbf{\begin{tabular}[c]{@{}c@{}}5.028\\ $\pm0.024$\end{tabular}}   & \textbf{\begin{tabular}[c]{@{}c@{}}0.598\\ $\pm0.002$\end{tabular}} \\ \hline
\end{tabular}
\caption{\textbf{PESQ and SDR improvements} compared to noisy audio baseline without the use of VoiceFilter. (Mean $\pm$ standard deviation.)}
\label{table:1} 
\end{table}

To evaluate the performance of different VF models, we use source to distortion ratio (SDR)~\cite{vincent2006performance} and perceptual evaluation of speech quality (PESQ)~\cite{rix2001perceptual} improvements. In both cases, a higher number indicates a better resemblance of the estimated speech to the clean speech of the reference speaker.

After conducting experiments of the individual task for three times, the mean results are shown in Table~\ref{table:1}. There are performance differences depending on which anchor vector is used. However, regardless of those differences, all proposed approaches show better performance compared to the conventional VF.
This is mostly because the conventional VF training aims to reconstruct clean spectrogram without considering the target speaker characteristic.
On the other hand, the proposed approach is able to effectively reconstruct the information of the target speaker even in a low SNR time-frequency block by using content information of the anchor speaker. 

Note that this improvement was achieved by only adding a loss function under the same VF structure, without any additional network parameter. 

\label{sec:exp}
\section{Conclusions}
\label{sec:con}
In this paper, we extended the conventional VoiceFilter system with additional training criteria, speaker representation loss and triplet SRL. Experimental results show performance improvement in speech separation via using the distance between speaker embedding spaces, \textit{i.e.} speaker content representation, to find the dominant attribute of the target speaker. This improvement was achieved without any additional network parameters or structures for the VoiceFilter system, but only with the application of our proposed SRL function. 

Our proposed loss is not just limited to source separation framework, but it could also be applied to other speech-related frameworks such as mask-based speech enhancement~\cite{chuang2019interspeech} and noise robust speaker verification~\cite{shon2019voiceid}. 

\vspace{12pt}\noindent\textbf{Acknowledgements.} We would like to thank Youna Ji, Minseok Choi, Bong-Jin Lee and Icksang Han for helpful comments.



\newpage
\label{sec:refs}

\bibliographystyle{IEEEbib}
\bibliography{shortstrings,mybib}

\end{document}